\documentclass[manuscript,authorversion, nonacm]{acmart}

\usepackage{hyperref}
\AtBeginDocument{%
  \providecommand\BibTeX{{%
    \normalfont B\kern-0.5em{\scshape i\kern-0.25em b}\kern-0.8em\TeX}}}

\acmConference[CHI '23]{ACM CHI Conference on Human Factors in Computing Systems}{April 23--28,  2023}{Hamburg, GER}
\begin{document}

\title{Policy Design in Data Economy: In Need for a Public Online News (Eco)System?}

\author{Viktoria Horn}
\email{viktoria.horn@uni-kassel.de}
\orcid{0002-2572-6575}
\author{Claude Draude}
\email{claude.draude@uni-kassel.de}
\affiliation{
  \institution{University of Kassel}
  \streetaddress{Pfannkuchstraße 1}
  \city{Kassel}
  \country{Germany}
}
%%
%%\author{Claude Draude}
%%\affiliation{%
%%  \institution{The Th{\o}rv{\"a}ld Group}
%%  \streetaddress{1 Th{\o}rv{\"a}ld Circle}
%%  \city{Hekla}
%%  \country{Iceland}}
%%\email{larst@affiliation.org}

%% \renewcommand{\shortauthors}{Trovato and Tobin, et al.}

\begin{abstract}
Socio-technical design embeds social investigations and inquiries into (Information) Technology Design processes. In this position paper, we propose, by using the aforementioned approach the design of technology and policies can simultaneously inform each other. Additionally we present data economy and particularly anchored online journalism platforms as use cases of policy need and design potentials.
\end{abstract}

%%\begin{CCSXML}
%%<ccs2012>
%%<concept>
%%<concept_id>10003456.10003462</concept_id>
%%<concept_desc>Social and professional topics~Computing / technology policy</concept_desc>
%%<concept_significance>500</concept_significance>
%%</concept>
%%</ccs2012>
%%\end{CCSXML}

%% \ccsdesc[500]{Social and professional topics~Computing / technology policy}

%% \keywords{Socio-technical Design, Data Economy, Online Journalism}

%% \received{20 February 2007}
%% \received[revised]{12 March 2009}
%% \received[accepted]{5 June 2009}

\maketitle

\section{Data Economy Need for IT Design Policy Regulation}

Algorithms are not merely seen as building blocks of technical products but likewise as building blocks of digital business models and even digital society as a whole, according to \citet{anderson2011deliberative} especially in today's society's information ecosystem.
Correspondingly \citet{gillespie2014relevance} developed the concept of \textit{Public Relevant Algorithms}, stating some kind of algorithms to be sufficient mediators of public knowledge and discourse. Those public relevant algorithms come with societal impact when shaping information ecosystems which are important for participation in public life, particularly in democratic processes \cite{gillespie2014relevance}. 
Navigating knowledge in algorithmically curated information systems leads to algorithms having an impact on the formation of opinion and on the creation of public spheres. 
On online platforms those public spheres are often embedded into techniques of digital data processing and economic utilization contexts, framed by the term ‘data-economy’ \cite{hess2019einfuhrung}.
This leads to the development of data-economic digital services that are embossed by creating user interfaces and algorithmic experiences designed to influence user’s decision-making abilities and nudge interactions that i.e. make users disclose more personal data than they normally would \cite{mathur2019dark, bosch2016tales}.
%% +societl challenges when considering display of information
Thus, not only privacy is at stake but also the design of user-friendly and socially desirable digital services. Claiming that the IT design of most digital services in data-economy is led by prevailing economic thinking, the need arises to design policies that account for societal demands.
% when designing technical artefacts in data economy, an exclusive design perspective is not sufficient, due to its embeddedness and effects on society. 

\section{Online Journalism Platforms in Data Economy: Democratic Values at Stake?}
Particularly ethical and democratic values of online journalism are seen as being threatened by the data economy ecosystem and its influence on IT development.
Potential threads are showing in phenomena like news recommender systems, that are primarly designed to fulfill economic purposes \cite{nechushtai2019kind} or native advertising, advertisement designed to look like journalistic content.
Societal problems that may arise by those technical design decisions are low exposure diversity of recommendations \cite{helberger2018exposure} and readers loosing faith in news media to be a credible source for democratic discourse \cite{bakshi2014and}.
Additionally, online news platforms have to compete with news aggregators of big tech companies like Google and Apple, emphasizing that market power and associated lock-ins play an important role in all ICT related domains \cite{breznitz2011value}. 
Thus, conflicts of interest between independent journalism, professional autonomy and economic resistance arise, accounting for methods and approaches that allow to design technology and ecosystem modalities jointly. 

\section{Socio-technical Design Approaches to mutually think about technical and societal design impacts and possibilities} 
Approaches that strive towards more inclusive, socially just, IT systems do not only focus on technological development but integrate societal, organizational and political perspectives as well. Most prominently, the socio-technical approach always considers social and technological aspects as interdependent and equally important in IT systems design. This means when planning a system, technical infrastructures, hard- and software, technical requirements along social structures, work tasks and people’s affordances need to be accounted for \cite{mumford2006story}. 
Furthermore, seeing societal issues as equally important would mean to factor in the sociopolitical context of a technological system and how its implementation affects different kinds of people \cite{draude2020situated, draude2022mapping}. Hence, democratic values such as non-discrimination, protection of marginalized people, self-determination for all are indispensable to consider in design policy recommendations. 
To realize self-determination, participatory design, reaching back to workers unions fights for workplace democracy in the 1970s in Scandinavia \cite{sundblad2011utopia}, should be considered. Participatory design seeks to increase user involvement in all stages of technological development but its impact also reaches beyond that. 
%Falls wir doch noch Platz brauchen kann der folgende Abschnitt raus
To encompass a systemic and political perspective on technology production, \citet{bodker2022what} formulate four strong commitments: “democracy at the workplace and beyond”; “empowerment of people through the process of design”; “emancipatory practices through mutual learning between designers and people”; “seeing people as skillful and resourceful in the development of their future practices”.
%Mögl. Kürzung Ende
While having a strong methodological base the real world application especially in times of global digital transformation is to be desired still. Participatory, socio-technical methods have been predominantly used for transformation processes on a smaller scale level, i.e. in individual organizations or in community projects.
%% Which work do we already contribute to a mutual adressing of technological and societal aims? 
%Ich würde BMBF herausnehmen bzw. hab ich schon, da das niemand versteht aber die Fördernummer und Institution bei acknowledgements angeben
A broader transferring of these approaches is something we strive for in our current research project FAIRDIENSTE\footnote{\url{https://www.uni-kassel.de/forschung/iteg/forschung/fairdienste}}. With interdisciplinary partners from academia, a public interest association and a German media house we look into different ways of fairly re-distributing value (economic and ethical) in digital service design. We address the mutual design of technology and organizational structures by rethinking business models as socio-technical intersections that should be designed participatively \cite{horn2022rethinking}. For this, we applied the participatory design approach to a co-design process of digital business model development. 
Through this, we managed to integrate socio-technical, participatory methodology and multiple stakeholders not just into the design of IT but into the company's organizational process of developing business models and technology in a co-creative fashion.  
What we learned from this could also be powerful to discuss and inform the designing of policies for IT design in the data economy, i.e. on the level of governmental decisions making. 
For example, matrix or systemic polices emphasizing education and (trans-)national competitiveness \cite{breznitz2011value} can be considered. According to \citet{helberger2018governing}, considering  
 "platforms and users as partners in regulation rather than as subjects" and relying on cooperative responsibility, which we support with our methodological contribution, is a more than promising approach.

\bibliographystyle{ACM-Reference-Format}
\bibliography{bibliography}

\end{document}